\documentclass{elsart}

\usepackage{amssymb}
\usepackage{amsmath}
\usepackage{graphicx}

\newcommand{\cf}{cf.\ }
\newcommand{\be}{\begin{equation}}
\newcommand{\ee}{\end{equation}}
\newcommand{\bea}{\begin{eqnarray}}
\newcommand{\eea}{\end{eqnarray}}
\newcommand{\Fig}[1]{Fig.\,\ref{#1}}
\newcommand{\Eq}[1]{Eq.\,(\ref{#1})}
\newcommand{\Eqs}[1]{Eqs.\,(\ref{#1})}
\newcommand{\la}{\langle}
\newcommand{\ra}{\rangle}
\newcommand{\nl}{\nonumber \\}

\newcommand{\Sec}[1]{Sec.\,\ref{#1}}
\newcommand{\Sch}{Schr\"{o}dinger\ }
\newcommand{\eph}{\emph{e-ph}\ }
\newcommand{\mbpar}[1]{\left( #1 \right)}
\newcommand{\mint}[2]{\int\limits_{#1}^{#2}}

\begin{document}
\begin{frontmatter}
\title{Conductance switching, hysteresis, and magnetoresistance in organic
     semiconductors}

\author[SDU]{J.~H.~Wei\corauthref{cor}},\ \ead{wjh@sdu.edu.cn}
\author[SDU]{S.~J.~Xie},\
\author[SDU]{L.~M.~Mei},\
\author[MPK]{J.~Berakdar},\
\author[HKUST]{YiJing Yan \corauthref{cor}}\ \ead{yyan@ust.hk}
\corauth[cor]{Corresponding authors.}
\address[SDU]{School of Physics and Microelectronics, Shandong
            University, Jinan 250100, China}
\address[MPK]{Max-Planck Institut f\"{u}r
            Mikrostrukturphysik, 06120 Halle, Germany}
\address[HKUST]{Department of Chemistry, Hong Kong University
           of Science and Technology, Kowloon, Hong Kong, China}

\date{\today}

\begin{abstract}
  The controllability of charge transport through an
organic molecular spin-valve system is theoretically investigated
on the basis of a Su-Schrieffer-Heeger model combined with the
non-equilibrium Green's function formalism. We show how the
formation of polaron in the organic sub-structure leads to a
hysteretic conductance switching, via sweeping either the bias
voltage or the electrochemical potential. We further obtain an
exponential dependence of the magnetoresistance as a function of
the applied bias voltage. The implications of calculated
 results in relation to experiments and device applications
 are addressed and commented.
\end{abstract}

\maketitle
\end{frontmatter}

\section{Introduction}
\label{thintro}

  Due to their compatibility in processing and
 tunability in electronic properties,
   $\pi$-conjugated organic semiconductors (OSEs) have been
 utilized broadly for various device
 applications \cite{Far01,Fri99121,Vos00442,Dat04S433,Roc05335}.
 Recent progress in the field of OSE electronics
 has revealed a number of technologically important
 transport characteristics, such as
 current rectification \cite{Met0055},
 Coulomb blockade \cite{And961323}, Kondo resonance \cite{Lia02725},
 negative differential resistance
 (NDR) \cite{Kra022927,Wal041229,Raw03377,Gui0455},
 bistable switching \cite{Col001172,Don012302,%
Ree013735,Li041949,Che05503,He017730,Blu05167},
 and giant magnetoresistance \cite{Ded02181,Xio04821}.

 An OSE system is characterized by not only
 its electronic structure but also its strong electron-phonon
 (\emph{e-ph}) coupling that induces
 polarons \cite{Hol59325,Hol59343}.
 In view of charging-induced conformational change
 \cite{Don012302,Sem003015}, the formation of nonequilibrium
 polaron should be a reasonable mechanism of the
 conductance switching and/or NDR in OSE electronics.
 This idea is first shown theoretically
 by Galperin et al.\ \cite{Gal05125} with a simplified model,
 in which the molecular wire consists
 of a single electronic state coupled with
 a single Brownian oscillator.

  Apparently, the Su-Schrieffer-Heeger (SSH) model \cite{Su802099}
 renders a more realistic description
 of an OSE system. It captures the essential
 characteristic of a conjugated molecule,
 where the strong \eph coupling leads it to the polaron
 charged states and dimerized ground state.
 Indeed, the SSH model has been
 remarkably successful in the study of
 the electronic conductivity and optical phenomena
 in $\pi$-conjugated OSEs.
 In addition to  conducting polymers, the SSH model has
 also been extended to
 molecular systems \cite{Tor00413,Nes01125422},
 carbon nanotubes \cite{Wei01207}, and DNA
 molecules \cite{Con004556}.
 We have recently \cite{Wei0682} exploited this model for
 an OSE system, and demonstrated that a NDR
 may appear due to the annihilation of a
 predoped polaron; i.e., an equilibrium charge transfer state
 that may exist before applying the external bias voltage.
 The resulting NDR can also be
 spin-polarization sensitive,
 leading to a magnetoresistance
 significantly larger than 100\%.

 In this work, we  elucidate the possibility of
 a polaron mechanism for the
 bistable switching and hysteresis in
 OSE nanostructures,  and further\
 for the giant magnetoresistance
 in OSE spin-valve devices where ferromagnetic
 metal electrodes are used.
  More importantly, we highlight that
 these technologically important properties
 are likely achievable by a wide range of OSE systems.
 To do that, moderate/typical values of
 the SSH parameters will be adopted in
 this work to avoid the
 formation of predoped polarons mentioned earlier.
  Besides the model and its parameters for the
 OSE spin-valve nanostructure of study,
 \Sec{ththeo} presents also the
 transport theory based on the Keldysh nonequilibrium Green's
 function (NEGF) formalism \cite{Kel641515,Dat95,Bra02165401},
  together with its implementation to the evaluation of
  current in response to linear sweeping bias voltage.
 Hysteretic conductance switchings
  controlled via bias voltage and via an electrochemical means
 are then numerically investigated
 in  \Sec{thswitchA} and \Sec{thswitchB}, respectively.
 The giant magnetoresistance property of OSE spintronic structures
 is addressed in \Sec{thgmr}.
 Finally, we conclude in \Sec{thsum}.

\section{Model and theory}
\label{ththeo}

\subsection{A  Su-Schrieffer-Heeger static polaron model}
\label{thmodel}

  We adopt a non-degenerate SSH Hamiltonian for
the isolate OSE system \cite{Ded02181,Con004556},
\bea \label{SSH}
    H_{\rm OSE}
  &=&
  \sum_{n,\sigma} \bigl\{
      \epsilon_o c_{n,\sigma}^+c_{n,\sigma}
     - [t_o-(-1)^nt_1
    -\alpha_o (u_{n+1}-u_n)]
\nl &\ & \quad \times
   (c_{n,\sigma}^+ c_{n+1,\sigma}+c_{n+1,\sigma}^+ c_{n,\sigma})\bigr\}
   +\frac{K_o}{2}\sum_n (u_{n+1}-u_n)^2 .
\eea
  Here, $c_{n,\sigma}^+$ ($c_{n,\sigma}$) denotes the creation
(annihilation) operator of an electron at the $n^{\rm th}$ site
 with spin $\sigma$ of the OSE system.
  The \eph coupling induced on-site lattice displacement
 parameters $\{u_n\}$ are treated classically
 and will be determined in a self-consistent manner
 to be described later (cf.\ \Sec{thNGF}).
 The input parameters involved in \Eq{SSH} are
the on-site energy $\epsilon_o$,
zero-displacement hopping integral $t_o$,
non-degeneracy parameter $t_1$,
electron-phonon coupling $\alpha_o$,
and lattice elastic constant $K_o$.

 For the electrodes, we choose two symmetrical $3d$
ferromagnetic (FM) transition metals; each of them is
described by a one-dimensional tight-binding
Hamiltonian with the Stoner-like exchange term \cite{Xie03125202},
\be\label{HF}
   H_{\rm FM} = \sum_{n,\sigma} \Bigl\{
      \epsilon_f d_{n,\sigma}^+d_{n,\sigma}
       +t_f(d_{n,\sigma}^+ d_{n+1,\sigma}
           +d_{n+1,\sigma}^+ d_{n,\sigma})\Bigr\}
  -\sum_n{J_f(d_{n,\uparrow}^+d_{n,\uparrow}
   -d_{n,\downarrow}^+d_{n,\downarrow})}.
\ee
Here, $d_{n,\sigma}^+$ ($d_{n,\sigma}$) denotes the creation
(annihilation) operator of a 3$d$-electron at the $n^{\rm th}$ site
with spin $\sigma$ of the metal electrode;
$\epsilon_f$ is the on-site energy of a metal atom,
$t_f$ the nearest neighbor transfer  integral,
and $J_f$ the Stoner-like exchange integral.

  The coupling between the OSE and each of the FM electrodes is
described by a tight-binding hopping integral,
\be \label{tc}
  t_c = \beta(t_f+t_o)  .
\ee
Here, $\beta$ is the OSE-FM binding parameter, which
for simplicity is set to be spin-independent.
However, charge transport can be spin dependent.
According to the two-current description,
the total current arises from the majority-spin
(up-spin) and the minority-spin (down-spin) of the
3$d$ electrons of the FM electrodes \cite{Fer691784}.
We also neglect the spin-flip processes during
the spin-dependent charge transport
that takes place between the FM electrodes.
  The above simple model for the molecule-metal coupling
allows us to focus on the effects of OSE molecular system
on the spin-dependent transport.

\subsection{Non-equilibrium Green's function formalism}
\label{thNGF}

  The quantum transport, in terms of
the conductance switching, hysteresis and
giant magnetoresistance behaviors of the model FM/OSE/FM system,
will be calculated by means of the
NEGF approach, based on the Keldysh formalism
\cite{Roc05335,Kel641515,Dat95,Bra02165401}.
 To proceed, the infinite FM/OSE/FM system is divided
into three distinct regions: The central scattering region (S)
that consists of the OSE together with a few metal atoms attached
to each of its ends \cite{Bra02165401}, and two semi-infinite FM
electrodes (L and R) that serve as charge
reservoirs and also set the temperature and electron
distribution of the junction in the steady state.
As results, the Hamiltonian for
the whole system is mapped onto  a $3\times 3$ block matrix,
\be\label{Htot}
 H_{\rm total} = \left(
\begin{array}{ccc}
  H_{\rm L} & H_{\rm LS} & 0 \\
  H_{\rm SL} & H_{\rm S} & H_{\rm SR} \\
  0 & H_{\rm RS} & H_{\rm R}
\end{array}
\right),
\ee
with $H_{\rm L}$, $H_{\rm S}$, and $H_{\rm R}$
being the isolated Hamiltonians for the L-electrode,
the central scattering S-region, and  the
R-electrode, respectively, while
$H_{\rm LS}=H_{\rm SL}^{\dag}$
and $H_{\rm RS}=H_{\rm SR}^{\dag}$
the couplings between them.

 The applied electric potential
bias $V$ does not change the electronic structure of
L and R leads, but just shifts their energy levels
by $eV/2$ and $-eV/2$, respectively.
This leads to the electron distribution function
$f(E,\mu_{\alpha})\equiv
[e^{(E-\mu_{\alpha})/k_{\rm B}T}+1]^{-1}$,
where $\mu_{\rm L/R}=\mu_{\rm eq}\pm eV/2$,
with $\mu_{\rm eq}=E_F$ being the Fermi energy of the FM electrode.
 In contrast to the electrodes, the effect of
applied bias on the S-region is characterized by
the electric potential function  $\Phi(x)$.
With the boundary conditions of $\Phi(x_1)=V/2$ and $\Phi(x_{N_{s}})=-V/2$,
it shifts the $n^{\rm th}$-site energy by $-e\Phi(x_n)$
in the S-region.
As it is correlated with the electronic and lattice structures of OSE,
$\Phi(x)$ should be evaluated together with the S-region
density matrix and lattice distortion in a self-consistent manner,
which will be elaborated soon [\cf \Eqs{DM}--(\ref{HMT})].

  Eliminating the L- and R-electrodes degrees of freedom
results in the NEGF of the central S-region \cite{Dat95},
\be \label{GR}
  G(E) = \frac{1}{E-H_{\rm S}+e\Phi(x)
   -\Sigma_{\rm L}(E)-\Sigma_{\rm R}(E)+i0_+}.
\ee
The effect of L- or R-electrode coupling is incorporated via
the self-energy,
\be \label{SigLR}
 \Sigma_{\alpha}(E)=H_{{\rm S}\alpha}
  \frac{1}{E -H_{\alpha}+i0_+} H_{\alpha{\rm S}} ; \ \ \alpha={\rm L, R}.
 \ee
In particular, $\Gamma_{\rm L/R}(E)\equiv i[\Sigma_{\rm L/R}(E)
  -\Sigma_{\rm R/L}^\dag(E)]$ amounts to
the electrode coupling-induced broadening function.

  The electron transmission probability function
reads in terms of the NEGF at a finite bias as \cite{Dat95}
\be \label{transmn}
  T(E,V) = \mbox{Tr} [\Gamma_{\rm L}(E) G(E) \Gamma_{\rm R}(E) G^\dag(E)].
\ee
The electric current can then be
evaluated as \cite{Dat95,Lan57223,Lan70863}
\be \label{I-V}
   I=\frac{2e}{h} \mint{-\infty}{\infty}
  T(E,V) [f(E,\mu_{\rm L})-f(E,\mu_{\rm R})]dE.
\ee
We shall show later that the above expression
remains practically valid in the evaluation
of current under the quasi-steady-state
voltage sweeping condition;
cf.~the comments to the end of this subsection.

 The NEGF determines also the density of states $D(E)$ and the
density matrix $\rho$ in the S-region \cite{Bra02165401}.
They are given respectively by
\be \label{DOS}
   D(E)=\frac{i}{2 \pi}\mbox{Tr}[G(E)-G^\dagger(E)],
\ee
and
\be  \label{DM}
  \rho= \frac{1}{2\pi}\sum_{\alpha={\rm L,R}} \mint{-\infty}{\infty}
   f(E,\mu_{\alpha})[G(E)\Gamma_{\alpha}(E)G^\dagger(E)]dE.
\ee
The number of carrier electrons in the S-region is
$N=\mbox{Tr}\mbpar{\rho S}$, with $S$ denoting the overlap integral matrix
that reduces to the unit matrix if
an orthogonal basis set is used.

 We are now in the position to complete the self-consistent
evaluation of the NEGF. The involving electric potential function
$\Phi(x)$ in \Eq{GR} is related to
the charge density, and the latter
depends further on the steady-state lattice distortion
at a given finite applied electric potential bias.

 Let us start with the Poisson equation that
relates $\Phi(x)$ and the on-site charge $\{\rho_{nn}\}$,
\be\label{Poisson}
  \nabla^2\Phi(x) = -\sum_n{\frac{\rho_{nn}}{\varepsilon_0}\delta(x_n-x)}.
\ee
 Here, $\varepsilon_0$ denotes the vacuum permittivity,
 and $x_n=(n-1)a + u_n$, with $a$ being
 the \eph coupling-free lattice constant and set to be 0.122\,nm.
 The \eph coupling-induced on-site lattice displacement
 is of the boundary conditions of $u_1=u_{N_o}=0$.
 To solve the
 Poisson equation we introduce
$\Phi(x)=(0.5-x/L)V+\delta\Phi(x)$,
 with $L$ denoting the S-region length.
 $\delta\Phi(x)$ satisfies the same
 Poisson equation, but with
 the boundary conditions
 of $\delta\Phi(x_1) =\delta\Phi(x_{N_o})=0$.
 It can be readily evaluated  in terms
 of the on-site charge $\{\rho_{nn}\}$ via the method
 of images, in which $\delta\Phi(x)$ is a sum of the
 individual image contributions from all point charges.
 To avoid the infinity associated with point charges,
 we follow the Appendix 3 of Ref.\,\cite{McL9113846} and
 analytically integrate the usual electrostatic potential within
 the small area around each point charge separately.

 Consider now the lattice distortion,
 which plays an important role in
 determining the properties of OSE
  due to the strong \eph interactions.
 For a steady or quasi-steady state,
 the lattice distortion is assumed to catch up with the variation
 of charges under a slow voltage sweep condition
 treated in this work. In this case,
 the OSE lattice distortion can be
 determined in terms of  the density matrix elements
 via the Hellman-Feynman variation
 theorem of $0=\partial{E_{\rm S}/\partial{u_n}
  = \partial{ {\rm Tr}(H_{\rm S}\rho)}}/\partial{u_n}$.
 It is that
\be\label{HMT}
 2\alpha_o(\rho_{n,n-1}-\rho_{n+1,n})+K_o(2u_n-u_{n-1}-u_{n+1})= 0.
\ee

  Equations (\ref{DM})--(\ref{HMT})
constitute a set of coupled equations
for solving $\rho$, $\Phi(x)$ and $\{u_n\}$
in a self-consistent manner by iteration.
 The converged result in the electric potential $\Phi(x)$ is
then used to determine the NEGF of \Eq{GR}, and
then the $I$-$V$ characteristics of \Eq{I-V}.
 The most time consuming part
 of the aforementioned self-consistent evaluation is \Eq{DM}.
 The involving integrand is
 non-zero over a broad energy range and often has
 sharp structures that  makes \Eq{DM} numerically challenging.
To facilitate this problem, let
\be \label{rhosum}
  \rho\equiv \rho_{\rm eq}+\delta \rho ,
\ee
where
\be \label{EDM}
  \rho_{\rm eq}= \frac{1}{2 \pi} \sum_{\alpha={\rm L,R}}
 \mint{-\infty}{\infty}
 f(E,\mu_{\rm eq})[G_{\rm eq}(E) \Gamma_\alpha(E) G_{\rm eq}^\dagger(E)] dE,
\ee
is the equilibrium density matrix, and
$\delta\rho$ the non-equilibrium correction.
 The evaluation of $\rho_{\rm eq}$
 requires a broad energy range and couples
 only with \Eq{HMT} as $V=0$ is implied.
 The involving integration will be tackled by taking the advantage of
 the contour integration technique \cite{Bra02165401,Wil825433}.
 On the other hand, the non-equilibrium
 correction $\delta\rho$ involves the self-consistent
 determination, together with \Eqs{Poisson} and (\ref{HMT}).
 It will be evaluated directly,
 as the integration is now required only for a limited energy range,
 where the integrands in \Eqs{DM} and (\ref{EDM})
 differ significantly.
 This is the same situation as the evaluation
 of current [\Eq{I-V}],
 and can thus be carried out with a fine grid.
 Note that $\rho_{\rm eq}$ is naturally real but
 $\delta \rho$ may not, unless a real basis set
 functions is employed. This is the case of this work,
 and the resulting $\rho$ does assume a real matrix.

 To conclude this subsection, let us comment
on the quasi-steady-state condition that will be applied
in this work for the evaluation of $I$-$V$ characteristics
subject to a sweeping voltage of $V(t)=V_0\pm V't$.
Here, $V_0 = 0$ or $V_{\rm max}$ for the sweeping up ($+$) or
down ($-$) direction, respectively;
$V'>0$ denotes the constant rate of bias sweeping,
which is typically $0.01\sim 0.1$ V/sec in experiments.
In our calculations of the quasi-steady-state
current of interest, we set $V'=0.1$ V/sec.
Note that all experimental observations
of the correlated electron and
lattice degrees of freedom
on the OSE  are functionals of density matrix $\rho$.
As the bias sweeping is linear, one can map the
time evolution of density matrix onto the bias sweeping one,
$\rho(t) \rightarrow \rho(V)$.
Starting from $\rho(t=0)$,
the density matrix evolution
is then evaluated via $\rho(t+\Delta t)=\rho(t)+\la \Delta \rho(t)\ra_{T}$,
where the second term denotes the time average
of $\Delta \rho(t)$ over the range of $[t,t+T]$.
with $T$ being the measurement duration (set to be $T=1$ second)
under the quasi-steady-state condition.
The corresponding $I(t)\rightarrow I(V)$
in each of the measurement time interval
is then formally the same as \Eq{I-V}.

\subsection{Model parameters}
\label{thpara}

  Reasonable values for the model parameters must be determined from
other sources, such as the band structure calculations and/or
experimental current-voltage ($I$-$V$) characteristics. In this
study, the FM electrodes are chosen to be cobalt metal.
With its experimental $3d$-band structure and
Fermi energy of $E_{\rm F}=-4.9$ eV
\cite{Xio04821,Pap86},
the tight-binding parameters for the Co metal in \Eq{HF} are determined
as $t_f=1.5$ eV, $J_f=1.45$ eV, and $\epsilon_f=-6.5$ eV.
The parameters of the OSE in \Eq{SSH} are chosen as
$\epsilon_o=-5.5$ eV, $t_o=2.5$ eV, $t_1=0.04$ eV,
$\alpha_o=5.0$ eV/\AA, and $K_o=21.0$ eV/\AA$^2$. These are the
modified values from those of a typical conducting polymer, but
lead to a relatively large band gap, which is about 2.8 eV for the
OSE length of $N_o=10$ sites considered in this work. In a
realistic OSE molecule, the on-site energy $\epsilon_o$ can be
 site-dependent and also can be adjusted by applying
 a gate voltage \cite{Dat04S433}.
The OSE-Co metal binding parameter $\beta$
affects the magnitude of current
sensitively, but it is hardly controlled due to the weak interaction.
In the calculations, one can adjust it so that the
calculated $I$-$V$ curve is of about the same amplitude of
the measured one.
We find that choosing $\beta=0.25$ for
the conductance switching via applied voltage sweep and
$\beta=0.5$ for that via electrochemical potential sweep will
produce the current in the magnitude of nA for some nano-size
molecules reported experimentally
\cite{Col001172,Don012302,Ree013735,Li041949,Che05503,He017730,Blu05167}.

 For a self-consistent evaluation of quantum transport based
on the NEGF formalism, the central scattering S-region is chosen
such that its Green's function can properly represent the bulk
property at the boundary. For the coupling parameters given above,
it is sufficient to have the S-region a total of $20+10+20$ sites;
i.e., to include 20 Co electrode atoms to each end of the model
OSE of length $N_o =10$. For the OSE-Co binding parameter used in
this work, it is found that the OSE subsegment is charge neutral
in the absence of applied bias ($V=0$). Changing the relative
energy levels of OSE with respect to electrodes may lead to
intra-regional charge transfer, forming a
preexisting (charged) polaron state in the OSE before applying bias
voltage \cite{Wei0682}.

\begin{figure}
 \includegraphics[width= \columnwidth]{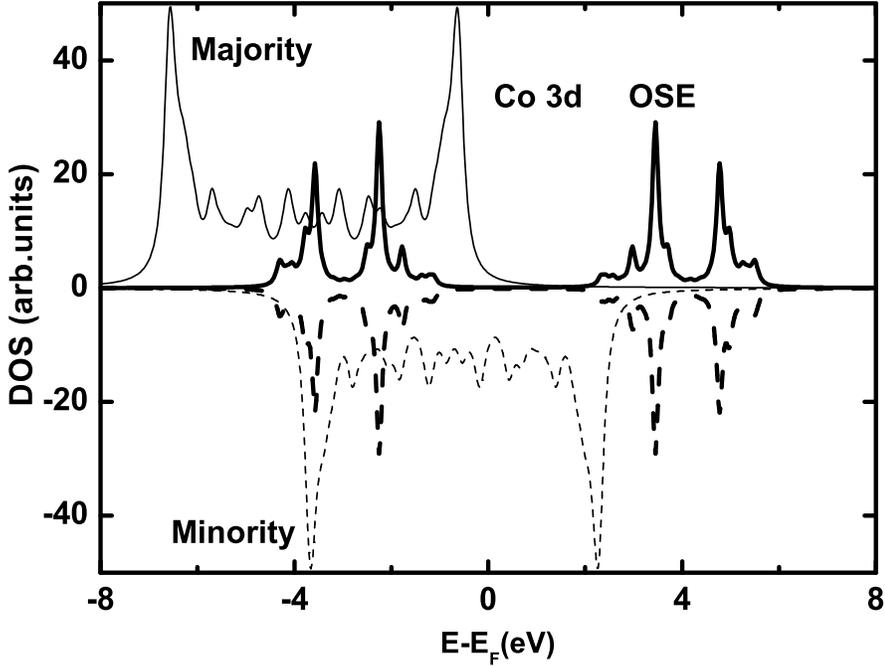}
 \caption{Density of states of Co $3d$ band and OSE structure at 11
  K. The solid (dashed) lines correspond to majority (minority)
  electronic bands. The thick (thin) curves are for OSE (Co),
  with the phenomenological Lorentzian width of 0.15 eV
  being included.}
\label{fig1}
\end{figure}

 Figure \ref{fig1} depicts the band structures
of the OSE and the Co metal electrodes at a low temperature
($T=11$ K). This figure summarizes the model parameters adopted in
this work (except for $\beta$ which in this figure is set to
zero). For the OSE, the evaluated band gap is about 2.8 eV, which
well reproduces the band diagram of the organic spin-valve system
in a recent experiment \cite{Xio04821}. For the Co electrodes,
our calculation is consistent with the real material \cite{Pap86},
showing a completely filled $3d$ band and a half-filled $3d$ band
for the majority-spin (up-spin) and the minority-spin (down-spin)
electrons, respectively.

 Quantitative simulations of experiment results
\cite{Xio04821,Yu05060408,Rud044898,Yu02201202} may need to
incorporate the spin dynamics, such as spin-flip, diffusion and
accumulation, which are however not included
 in this work. Despite of its simplicity,  we shall show
soon that the present model possesses
some interesting properties that can highlight the
important transport mechanism in relation to experiments.

\section{Hysteretic conductance switching via bias voltage sweep}
\label{thswitchA}

  In this section we report the hysteretic bistable switching $I$-$V$
behavior of the model Co/OSE/Co structure.
In contact with experimental reality \cite{Col001172,He017730},
the bias sweep rate in our calculations is fixed at
a representing value of 0.1 V/sec, at which
the quasi-steady-state condition is
reasonably satisfied.

 There are two distinct transport measurement configurations,
parallel (P) and antiparallel (AP), with respect to the relative
magnetization orientation of two FM electrodes.
Generally, in the
two-current model, both the majority-majority and
minority-minority (or majority-minority and minority-majority)
transports are permitted in the P (or AP) configuration.
However, for the cobalt electrodes in the present work,
the full-filled up-spin electrons cannot transport
in P-configuration under a positive bias voltage $V>0$
and the charge
carriers are only the half-filled down-spin electrons.
Contrary, the charge carriers in AP-configuration under
$V>0$ are only the up-spin electrons, driven from the
full-filled majority-spin
band of L-electrode to the half-filled minority-spin band of
R-electrode of opposite magnetization orientation.
\begin{figure}
\includegraphics[width= \columnwidth]{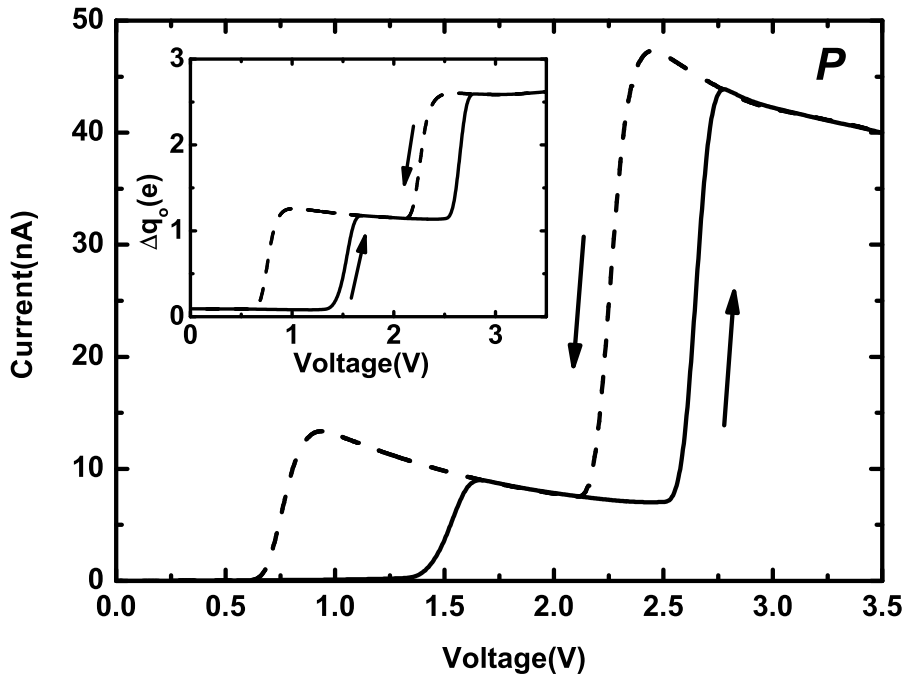}
\caption{The hysteretic current as the function of sweeping bias
potential
 for the model OSE in junction with Co electrodes in
 the parallel configuration.
 The linear voltage sweep rate is at 0.1 V/sec.
 The insert is the bias voltage-induced charge variation
 in  the OSE substructure. The
 OSE-Co metal binding parameter $\beta=0.25$.
 } \label{fig2}
\end{figure}

\begin{figure}
\includegraphics[width= \columnwidth]{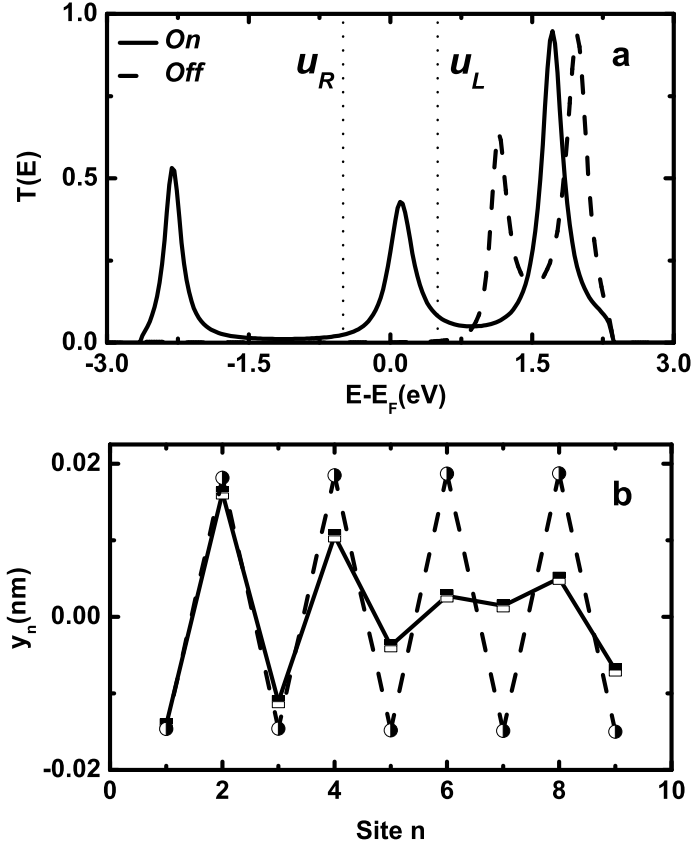}
\caption{Bistable switched-on (solid-curves) and switched-off
(dashed-curves)
 states for the P-configuration
 at bias voltage $V=1.0$ V. (a) Transmission coefficient of
 down spin electrons. The vertical dot lines indicate the window
 between $\mu_L$ and $\mu_R$ for the charge transmission;
 (b) Lattice distortion $y_n\equiv u_{n+1}-u_n$
 of the OSE substructure.
} \label{fig3}
\end{figure}

 Figure \ref{fig2} depicts the resulting hysteretic $I$-$V$
characteristic in P-configuration.
The bias sweep directions are indicated with the arrows.
The OSE nanostructure is shown to have
an insulating state and two conducting states
in the indicated sweeping bias range.
These conducting states are switched on (off), respectively,
at $1.5\pm 0.2$\,V ($0.75\pm 0.2$\,V) and $2.6\pm 0.1$\,V
($2.1\pm 0.1$\,V) in the sweep up (down) direction.
 The voltage-triggered two-step hysteretic
 conductance switching has been observed
 experimentally in some molecular
 nanojunction systems \cite{He017730}, often associating
 with voltage-triggered molecular confirmation
 change.
 On the other hand, that observed in \Fig{fig2},
 according to the \eph coupling
 Hamiltonian used here (\cf \Sec{ththeo}),
 can only be of polaron in nature.
 The insert in \Fig{fig2} shows the sweeping bias-induced
 charge modification $\Delta q_o$ within the OSE sub-structure.
 The synchronization between the $\Delta q_o$-$V$
 and $I$-$V$ curves demonstrates clearly that the two
 voltage-triggered conducting states
 have 1.15e and 1.44e charges, respectively.
 These are negatively charged polaron states,
 with the clear evidence of the \eph coupling induced
 lattice distortion being presented soon (\cf \Fig{fig3}).

  To further elaborate what happens to
 the conductance bistability of the OSE nanostructure,
 \Fig{fig3} depicts the transmission function
 and lattice distortion for the bistable states,
 exemplified with the P-configuration at $V=1$\,eV (\cf \Fig{fig2}).
 The resulting window between $\mu_{\rm L}$ and $\mu_{\rm R}$
 for the electric transmission $T(E)$, as indicated in \Fig{fig3}a,
 clearly distinguishes between the conducting state (solid curve)
 and the insulation state (dash curve).
 Note that the lattice
 distortions in \Fig{fig3}b are reported
 in terms of the \eph coupling induced
 bond length variation $y_n=u_{n+1}-u_n$ between
 nearest neighbors. The dashed curve there
 for the insulating state is effectively the same
 as the equilibrium ($V=0$) lattice distortion.
 In contrast, the solid curve in \Fig{fig3}
 demonstrates that a charged polaron state is formed
 in the conducting state toward the right-side of
 the OSE-substructure due to the external bias-induced spatial
 symmetry broken.

  The polaron mechanism for the observed hysteretic conduction
 switching behavior in \Fig{fig2}
 can now be summarized as follows.
 In the equilibrium state ($V=0$), the OSE substructure
 is charge neutral, with its lowest-unoccupied and
 highest-occupied molecular orbital
 (LUMO and HOMO) electronic wave functions delocalized
 over the entire S-region.
 The chemical potentials of electrodes are the same as the Fermi energy,
 $\mu_{\rm L}=\mu_{\rm R}= E_F$, locating in between
 the LUMO and HOMO levels of the OSE; no current is observed
 and the OSE is in an insulating state.
 Applying a positive bias $V>0$, which
 elevates $\mu_{\rm L}=E_F+eV/2$ and
 depresses $\mu_{\rm R}=E_F-eV/2$, drives electron
 transport from L-electrode to R-electrode through
 the unoccupied levels of OSE substructure.
 Since the two bistable conductance states in \Fig{fig2} are of
 the same physical origin,
 let us focus on the first one for illustration.
 As the sweep-up voltage approaches to the
 vicinity of the LUMO level,
 electron starts to migrate into the OSE, followed
 by the \eph coupling-induced reorganization
 that leads to the formation of polaron.
 In other words, $V_{\rm on}=2(E_{\rm LUMO}-E_F)/e$ defines
 the switch-on voltage of the first conducting state.
 However, as the OSE gets charged,
 the \eph coupling-induced reorganization
 process occurs, forming a polaronic gap state that
 serves as the first conducting state in \Fig{fig2}.
 The fact that $E_{\rm polaron}<E_{\rm LUMO}$ due to the reorganization
 accounts therefore for the observed hysteresis,
 since $V_{\rm off}=2(E_{\rm polaron}-E_F)/e$ is
 the switch-off voltage in the down-sweep direction.
 In the case of study here,
  $V_{\rm on} \approx 1.5$\,V  and $V_{\rm off} \approx 0.75$\, V
 (\cf \Fig{fig2}). The reorganization energy of
 the negatively charged polaron gap state
 can be obtained as $E_{\rm LUMO}-E_{\rm polaron}=
 e(V_{\rm on} - V_{\rm off})/2 \approx 0.37$\,eV.

\begin{figure}
\includegraphics[width= \columnwidth]{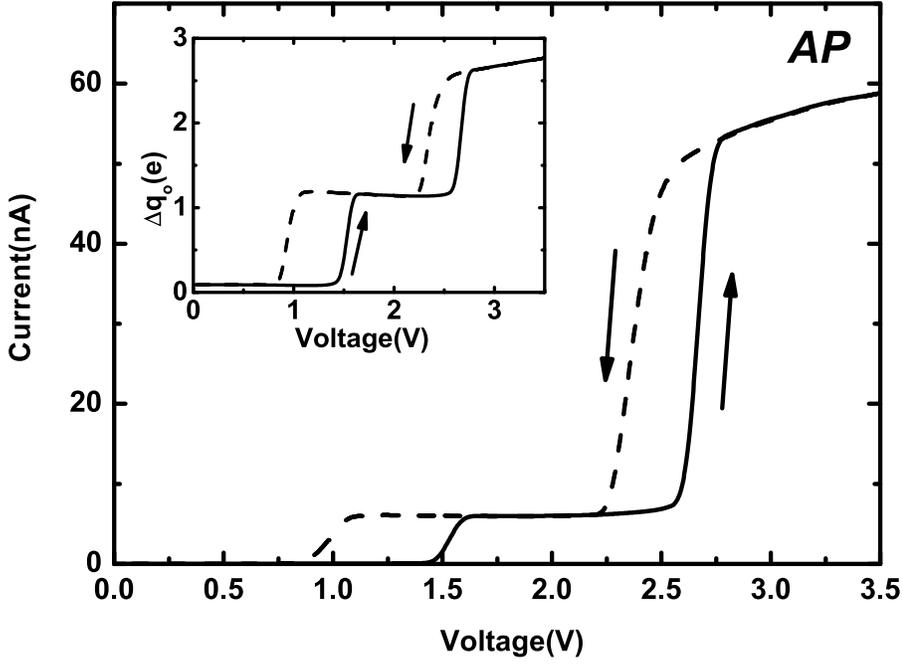}
\caption{ Same as \Fig{fig2} but for the
  antiparallel configuration of the Co electrodes.
} \label{fig4}
\end{figure}

 Figure \ref{fig4} depicts the hysteretic $I$-$V$ characteristics
 and the corresponding charge modifications for the
 AP-configuration of the electrode magnetization directions.
 Similar to the P-configuration, the two-step
 hysteretic conductance switching behavior
 appears also here.
 This observation indicates that the hysteresis itself
 is originated from the OSE due to the polaron formation,
 regardless the contact is magnetic and nonmagnetic
 in this aspect.
 However, in the present Co/OSE/Co model study,
 the hysteresis curves in the AP-configuration are narrower
 than those in the P-configuration.
 This difference gives us an additional handle by
 using a magnetic contact on the conductance control
 since one can alter from the P- to the AP-configuration
 of magnetization by applying an
 external magnetic field, especially when the electrode
 is made of soft magnetic material.

\section{Hysteretic conductance switching via electrochemical potential
 sweep}
\label{thswitchB}

 We now turn to another type of conductance switching,
controlled by sweeping the Fermi energy [or $\epsilon_f$
effectively in eq (2)] of both electrodes simultaneously via an
electrochemical means, or equivalently sweeping the gate voltage
via an microelectronic means that adjusts the on-site energy of
the OSE.
 Here, the bias voltage is kept fixed at a small
 value to open a transmission window between $\mu_{\rm L}$ and
$\mu_{\rm R}$.  For instance, in the experiment of
 He et al.~\cite{He017730}, the current through
Au-polyaniline-Au nanoelectronic junctions
 was measured under a linear sweep of the electrochemical potential
 at a fixed bias voltage of 20 mV.
 The fast conductance switching behaviors of the
 electrochemical system is closely related to chemical and biosensor
applications \cite{He017730,Sol03296}, while the gate voltage
modulation is of great importance for the organic transistors and
memories \cite{Li041949,Xu052386}. They share the same mechanism;
i.e.\ the control of the polaron (reduced/oxidized) state of the
molecule by electrochemical potential or gate voltage,
and can therefore be qualitatively illustrated
with the same model.
  As discussed earlier, the polaron formation
 is responsible to the conductance switching.
 It arises from the coupling between
 the electronic states of the OSE near the LUMO energy
 and that of the metal leads near $\mu_{\alpha}$ ($\alpha={\rm L,R}$).
 When $\mu_{\alpha}$ approaches to the
 polaron energy level (which is lower than the LUMO
 of the ground state) an abrupt change of current occurs.

  In the following calculations, we fix
 the electrochemical potential sweeping rate at 0.1 eV/sec
 that supports the quasi-steady-state condition.
 As described to the end of \Sec{thNGF},
 \Eq{I-V} together with \Eqs{DM}--(\ref{HMT})
 are used again for the evaluation of the
  quasi-steady-state current as the
 time-averaged one in the interval of every second.

\begin{figure}
\includegraphics[width= \columnwidth]{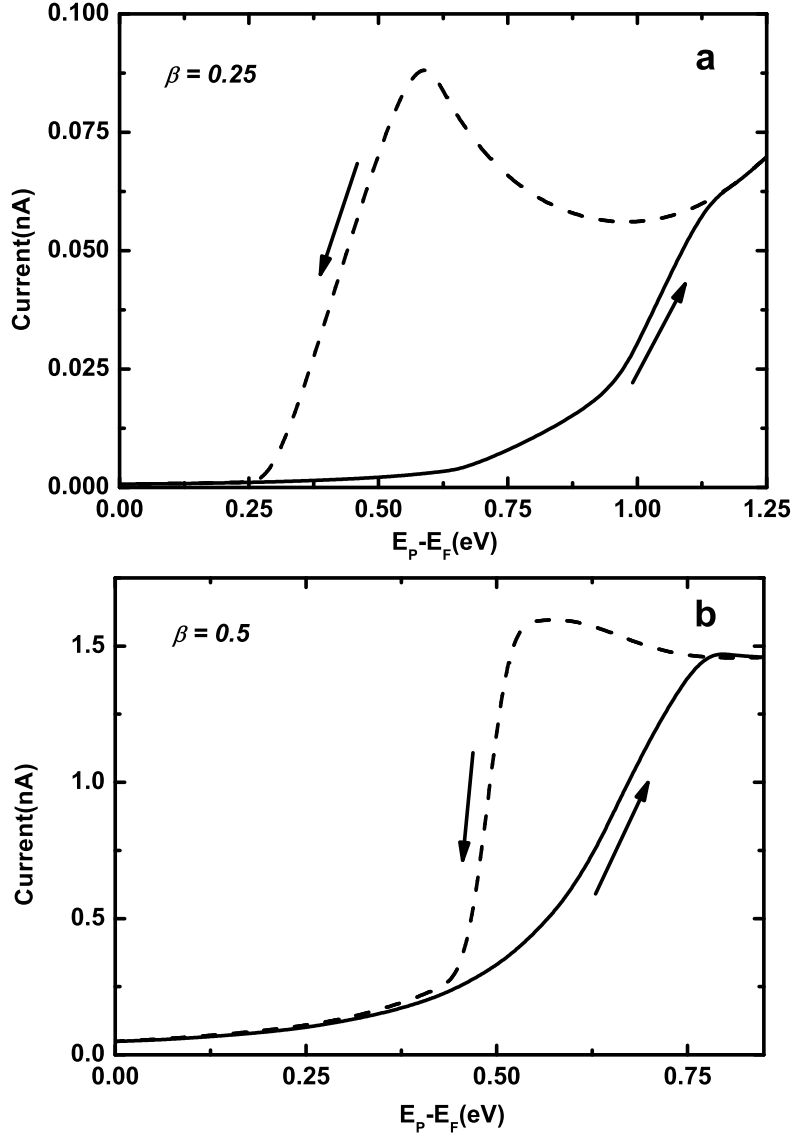}
\caption{The hysteretic current as the function of sweeping
electrochemical
  potential for the model OSE in junction with Co electrodes in
 the parallel configuration, at the fixed bias voltage of $V=20\,$mV.
 The OSE-Co binding parameter (a) $\beta=0.25$ and (b) $\beta=0.5$,
 respectively. The initial Fermi level $E_F=4.9$ eV.
} \label{fig5}
\end{figure}

  Figure \ref{fig5} shows the dependence
 of the charge transport current on a sweeping of
 the electrochemical potential $E_P$, in terms of
 its difference from the initial Fermi level $E_F$ ($=4.9$ eV) of the metal leads
 in the P-configuration of electrodes.
 The bias voltage between the electrodes is kept at
 a constant value of 20 mV.
 The OSE-Co metal binding
 parameter is $\beta=0.25$ in \Fig{fig5}a
 and $\beta=0.5$ in \Fig{fig5}b,
 respectively.
 While the value of $\beta$ affects the magnitude of current,
 the hysteretic behavior of conductance switching
 is shown to be an intrinsic character of the OSE,
 regardless the details of OSE-electrode coupling.
 Making in contact with the experimental output of current, which
 is of nA in magnitude for a nano-sized molecule \cite{Li041949,He017730},
 $\beta=0.5$ is favored for the calculations of the
 potential-dependent conductance switching behavior.

 As shown in \Fig{fig5}b, enhancing the potential leads to
 a rapid increase of the current during
 $E_P-E_F \sim 0.5$ eV and 0.75 eV,
 as the OSE switches from the insulating to the conducting state.
 The hysteresis behavior here is attributed to the
 formation of a {\it bipolaron} state that localizes two electrons in
 the OSE region. The formation of a bipolaron (not depicted here)
 is further confirmed with various OSE chain lengths of
  $N_o=10\sim 40$.
 The electrochemical potential-dependent charge
 transport behavior is similar to its bias voltage-dependent
 counterpart. In both cases, the same physical
  mechanism; i.e., the polaron-assisted
  charge transport is operational.
 A similar phenomenon is also
 found for the AP-configuration of electrodes.

  The potential dependent spin-polarized charge
 transport through OSE
 has been studied theoretically by
 Xie \emph{et al.} \cite{Xie03125202} in
 the context of the ground-state properties
 of ferromagnetic metal/conjugated polymer interface.
 The conclusions there are that
 electrons can transfer to the polymer
 and form bipolarons from the magnetic layer
 through an interfacial coupling by adjusting the relative chemical
 potentials of the contact and the polymer. This conjecture is
 confirmed by the present calculations.

\section{Magnetoresistance effects}  
\label{thgmr}

\begin{figure}
\includegraphics[width= \columnwidth]{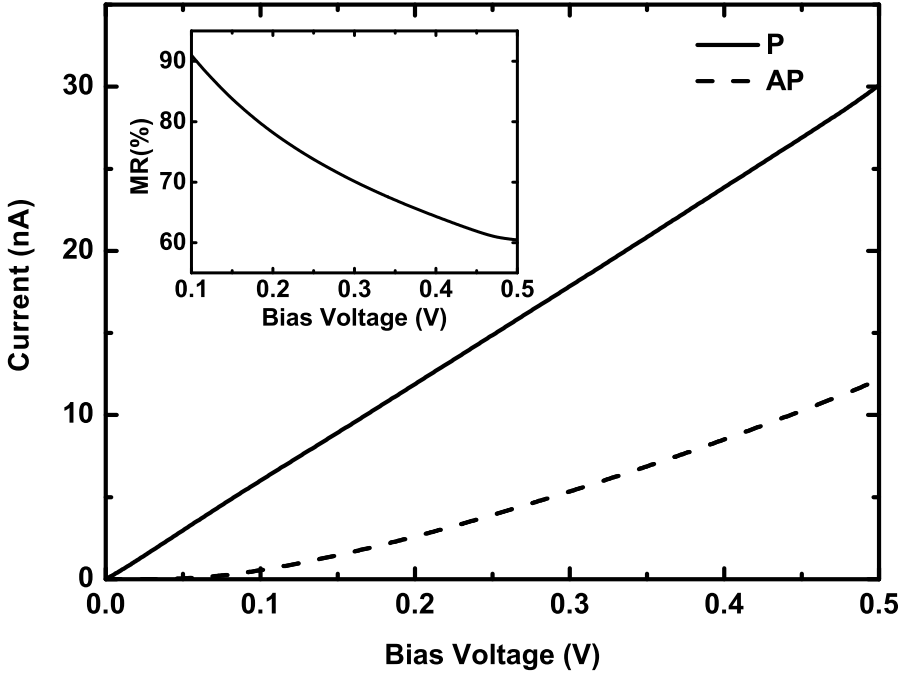}
\caption{The $I$-$V$ characteristic of the OSE spintronics at the
 conducting state, with the electrochemical potential
 of $E_P = E_F + 0.8\,$eV,
 in the parallel (solid curve) and
 antiparallel (dashed curve) configurations.
 The insert is the MR as the function of bias voltage.
}\label{fig6}
\end{figure}

  Now we address the magnetoresistance property of the model
 Co/OSE/Co spin-valve system,  measured as
 \be\label{GMR}
  \text{MR} = \frac{R_{\rm AP}-R_{\rm P}}{R_{\rm AP}}.
 \ee
Here, $R_{\rm AP}$ ($R_{\rm P}$) denotes the
electric resistance with the antiparrellel  (parallel)
magnetization orientation of two FM electrodes.
To study that, we turn on the conductance of OSE
by choosing $E_P=E_F+0.8$ eV (\cf \Fig{fig5}),
and calculate the steady-state currents
[\Eq{I-V} with \Eqs{DM}--(\ref{HMT})].
The resulting $I$-$V$ characteristics
of the model Co/OSE/Co spin-valve device
in the conducting state for both the P- and AP-configurations
of electrodes are shown in \Fig{fig6},
and the corresponding MR behavior in the insert.
The $I$-$V$ characteristics for the P-configuration
exhibits  roughly a linear behavior
(with $R_{\rm P}\approx 17\,$M$\Omega$)
and can be considered as a metallic phase.
In contrast,  for the  AP-configuration the
current increases much slower
and is nonlinear as the bias voltage increases.
The resulting MR shown in the insert of \Fig{fig6}
decays nearly exponentially as the applied
bias voltage increases from $V=0$ where MR$\,\approx\,$100\%
to $V=0.5\,$V where MR$\,=\,$60\%.
The relatively large MR here in comparing with
the typical experimental values may partially be attributed to
the fact that our model system does not include spin-flip processes.
 As shown in \Fig{fig6}, $R_{\rm P}$ is almost constant
 for the range of voltage reported here;
 thus, the bias-dependence of MR is governed by
 the $R_{\rm AP}$ in the AP-configuration,
 where the charge carriers in $V>0$ are
 the up-spin electrons driven from the full-filled
 (majority) band of the L-electrode to the half-filled
 (minority) band of the R-electrode (\cf \Fig{fig1}).
 As the up-spin electrons traverse
 the polaron level of the OSE barrier,
 the electric resistance $R_{\rm AP}$ decreases
progressively as the bias voltage increases,
resulting in the MR of an appearance of exponential decay.

\begin{figure}
\includegraphics[width= \columnwidth]{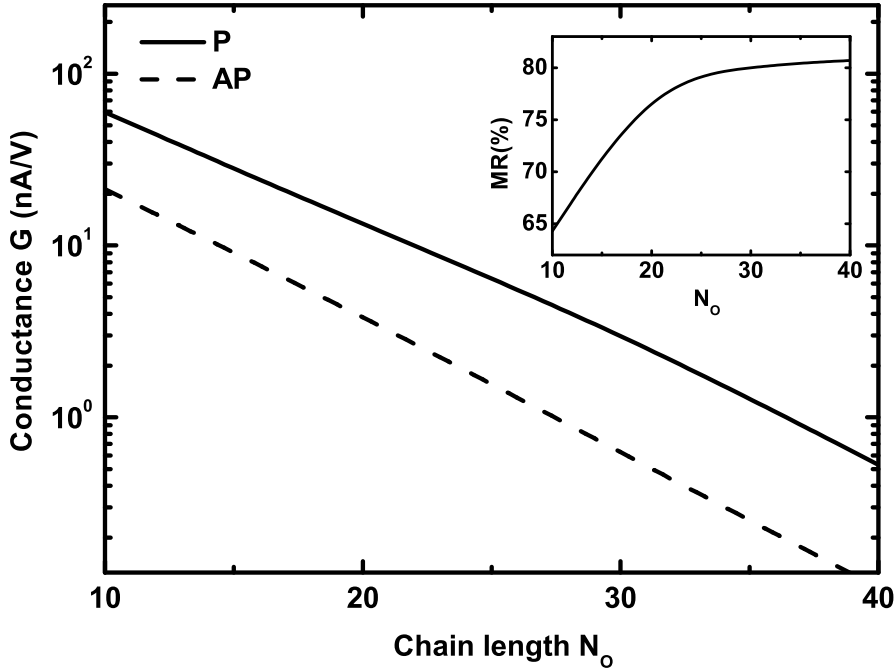}
\caption{The length dependence of the conductance in
  the parallel (solid curve) and antiparallel (dashed curve)
  configurations, at the bias voltage of $V=0.4\,$V.
 The insert is the corresponding length dependence of the MR.
} \label{fig7}
\end{figure}

 We now turn to the effect of OSE length on the transport,
which has been considered so far only with the fixed value of $N_o=10$.
  The experimental conductance often decreases
exponentially as the length of the
molecular chain increases \cite{Wak0319}.
The length-dependent conductance
can be investigated theoretically with
the help of the equilibrium Green's function formalism
\cite{Asa05085431}, and the exponential
length dependence of the conductance
characterizes the coherent transport \cite{Yan02225}.
 Figure \ref{fig7} depicts our theoretical results using
the NEGF formalism at the bias voltage $V=0.4$ V.
The evaluated conductances do show near exponential dependences,
for both the P- and the AP-configuration. The latter
decays faster, leading to the non-uniform OSE-length
dependence of MR, as shown in the insert of \Fig{fig7}.

\section{Summary and concluding remarks}
\label{thsum}
  In summary, we have investigated
 the spin-dependent charge transport in  organic spintronics  devices,
 on the basis of the SSH model combined with the
 NEGF formalism.
  We have demonstrated the polaron mechanism for the conductance
 switching, which can be controlled either by an appropriate
  bias voltage (\Sec{thswitchA}) or by changing the
 electrode electrochemical potential (\Sec{thswitchB}).
  Despite of its dependence on the electrode property,
 the conductance switching is generally a property
 of nanosize organic semiconductors.
  To have a one- or two-step conductance switching,
 it is not necessary for the lead to be magnetic.
  It is believed that the two-step conductance switching can
 be observed even in some nanostructures with nonmagnetic
 or paramagnetic electrodes. However, using
 magnetic electrodes does offer an additional
 handle on the conductance control (see the
 discussion to the end of \Sec{thswitchA});
 thus it has potential applications in organic electronics.
 We have also studied the behavior of magnetoresistance (\Sec{thgmr}).
 The calculated magnetoresistance shows an exponential decay
 as the applied bias voltage increases in the $V>0$ region.
 The dependence of the magnetoresistance
 on the length of molecular junction is also addressed.

 It is worth to mention here that a decay behavior of
 MR has been reported experimentally
 in an LSMO/OSE(Alq$_3$)/Co  spin-valve device,
 where the MR has a maximal value at $V\approx 0$
 and decreases exponentially as the applied
 voltage increases \cite{Xio04821}.
 Similar decay behavior is also
 observed  for OSEs with magnetic
  tunnel junctions with LSMO and Co or
 CoFe electrodes \cite{Ter994288,Hay028792}.
 These experiments are for layer devices with the OSE
 thicknesses that may lead to the
 spin-dependent transport of diffusive.
 The spin diffusion can be incorporated
 phenomenologically into the present model
 to account for the relatively fast decay
 behavior observed experimentally.
 It is however beyond the scope of this paper
 and will be considered elsewhere.

  The present calculations are done for the steady or quasi-steady state
 electric transport.  The dynamical polaron effects
 that are not considered in this work
 may be importance for the transport in organic spintronics,
 especially in the presence of a time-dependent external field.
  Some theoretical investigations have been devoted to this issue.
 The method used there is the wave-packet
 evolution approach \cite{Yu9916001} or
 non-adiabatic time-dependent \Sch equation
 method \cite{Wu03125416}.
 The resulting transport mechanism depends rather sensitively
 on the time scale of the electronic motion than
 that of the lattice distortion.
 Increasing the lattice vibrational frequency, the
 transport can change from electron-like to polaron-like
 behavior. The strong lattice fluctuation may also lead to
 the polaron transport mechanism even in
 the low phonon-frequency regime \cite{Yu9916001}.
 Nevertheless, the polaron mechanism studied in this work
 for the hysteretic conductance switching and giant magnetoresistance
 is expected to remain at least qualitatively valid.

\section*{Acknowledgment}
  Support from the National Natural Science Foundation of China
(No.\ 10474056 and No.\ 10234010) and the
 Research Grants Council of the Hong Kong Government
 (No.\ 605105) is gratefully acknowledged.

\clearpage

\end{document}